\documentclass[useAMS,usenatbib,usegraphicx]{mn2e}

\title{Astrometric signal profile fitting for Gaia}

\author[M. Gai, R. Cancelliere and D. Busonero]{M. Gai$^{1}$ 
\thanks { E-mail: gai@oato.inaf.it (MG); cancelli@di.unito.it (RC); 
busonero@oato.inaf.it (DB)}, 
R. Cancelliere$^{2}$ and D. Busonero$^{1}$
\\ 
$^{1}$Istituto Nazionale di Astrofisica - Osservatorio Astronomico di 
Torino, V. Osservatorio 20, 10025 Pino T.se (TO), Italy \\ 
$^{2}$Dipartimento di Informatica, Universit\`a di Torino, 
C.so Svizzera 185, 10149 Torino, Italy }

\begin{document}

\date{}

\pagerange{\pageref{firstpage}--\pageref{lastpage}} \pubyear{2002}

\maketitle

\label{firstpage}

\begin{abstract}
A tool for representation of the one-dimensional astrometric signal
of Gaia is described and investigated in terms of fit discrepancy
and astrometric performance with respect to number of parameters required.
The proposed basis function is based on the aberration free response
of the ideal telescope and its derivatives, weighted by the source
spectral distribution. The influence of relative position of the detector
pixel array with respect to the optical image is analysed, as well
as the variation induced by the source spectral emission. The number
of parameters required for micro-arcsec level consistency of the reconstructed
function with the detected signal is found to be 11. Some considerations
are devoted to the issue of calibration of the instrument
response representation, taking into account the relevant aspects
of source spectrum and focal plane sampling. Additional investigations
and other applications are also suggested. 
\end{abstract}

\begin{keywords} 
astrometry -- methods: numerical -- instrumentation:miscellaneous. 
\end{keywords}

\section*{Introduction }

In the framework of the data reduction for Gaia \citep{2005ASPC..338....3P,2009IAU...261.1601L},
the issue of a convenient representation of the instrument response,
i.e. of the detected signal profile, at the micro-arcsec 
(hereafter, $\mu as$) level, is crucial to science data modelling,
calibration and analysis. 

Since a large fraction of the astrometric data of Gaia is one-dimensional,
obtained by across scan binning during the CCD readout with the purpose
of reducing the sheer amount of data, the investigation and analysis
is referred to single-valued functions of one variable, i.e. intensity
vs. focal plane position. The signal coordinate is basically coincident
with the high resolution direction of the telescope, and with the
scanning direction of the satellite. The one-dimensional signal is
referred to as Line Spread Function (LSF) in the following, for similarity
with the optical signal of an infinite slit, although the term is
only applicable in a loose sense: e.g., the signal from one
source may suffer contamination by other sources at some distance
in the across scan direction, which would not be the case for real
LSFs. Also, the finite readout area implies a small variation of the
detected photon fraction with the across scan position on the detector,
which would not happen for an LSF of negligible across size.

The signal profile from a real instrument differs from the ideal telescope
response because of optical aberrations, instrument operation, detector
characteristics, and a number of environmental aspects influencing
them; also, the signal depends on the individual source spectrum.
The detected signal can evolve during the mission lifetime due to
degradation of both optical and electronic components. 
In the analysis described in this document, the case of comparably 
small perturbation to the ideal image, represented by small optical 
aberrations, is dealt with; the signal model can be extended to larger 
image degradation with straightforward
modifications, but the precision can be expected to suffer progressive
degradation as well. The modeling precision could then be retained
by a description based on a larger number of parameters. 

The proposed modeling framework is based on a set of functions, described
in Sec. \ref{sec:Signal_model}, derived from the monochromatic, aberration
free LSF of an idealised telescope retaining the basic geometric characteristics
of the Gaia instrument, i.e. the rectangular aperture width $L_{\xi}=1.45\, m$.
The source spectrum is explicitly inserted in the construction of
the polychromatic LSF, again referred to the aberration free telescope.
Some of the additional image degradation effects associated with detector
characteristics and operation are then introduced (Sec. \ref{sub:Detection-effects}).
The signal degradation effects associated to a realistic instrument
and operation, including known effects not explicitly included in
the model, are then described by the expansion of individual signals
in terms of the proposed function set.

\begin{figure*}
\includegraphics[scale=0.7]{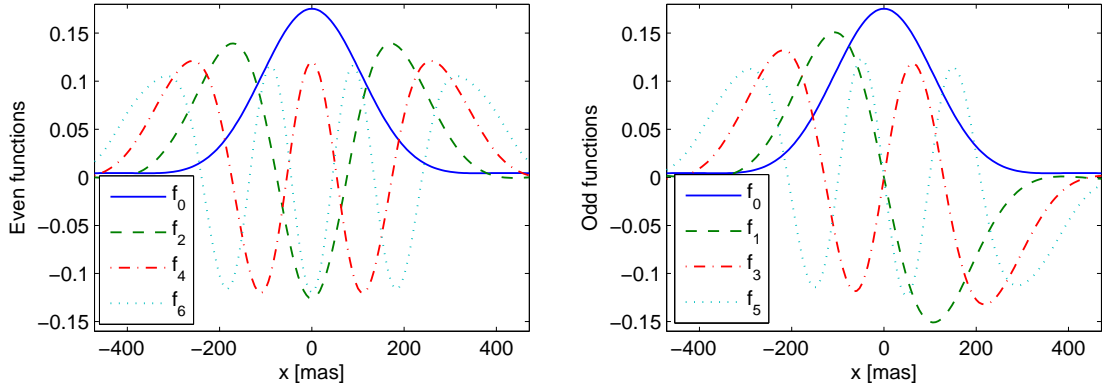} 
\caption{\label{fig:Polychromatic-detected-LSF}Polychromatic basis functions
for a 6000 K source: even terms (left) and odd terms (right)}
\end{figure*}

The implications of the proposed model are evaluated in Sec. \ref{sec:Simulation}
by simulation over a range of aberrations, of sampling offset (meaning
the relative position of the optical image vs. the detector pixel
array), and of source spectral type, modeled as blackbodies at different
effective temperatures. The simulation is implemented in the Matlab
framework. 
{\em Remark: 
it is assumed that variation of relevant system parameters (e.g.
detector electro-optical characteristics) can be represented, at 
first order, by optical aberrations inducing a similar effect on the 
detected signal. } 

The number of parameters required for proper fit of the detected signal
with adequate precision are derived, and some of the mathematical
and physical characteristics are discussed. In Sec. \ref{sec:Calibration}
the calibration of the model parameters is addressed with respect
to some measurement aspects. In Sec. \ref{sec:Discussion} we discuss
some of the possible developments related to the proposed model, in
terms of improvement of its performance and generality, as well as
usage in other cases, like the photometric and spectroscopic instruments
of Gaia, or other astronomical equipment. Finally, we draw our conclusions
on the effectiveness of the proposed signal expansion model.

\section{Signal model}

\label{sec:Signal_model} The starting point of our derivation is
the monochromatic response of an ideal instrument, i.e. the signal 
generated by 
an infinite slit, without aberrations. This is the well
known squared sinc function, hereafter called \emph{parent function}
\citep{1999prop.book.....B}, depending on an adimensional argument
related to the focal plane coordinate $x$, the wavelength $\lambda$
and the aperture width $L_{\xi}$, as 
\begin{equation}
f_{0}^{m}\left(x\right) = \left[\frac{\sin\rho}{\rho}\right]^{2},\;\rho 
= \pi\frac{xL_{\xi}}{\lambda F}\,. 
\label{eq:Parent_function} 
\end{equation}
 Many function families known in mathematics are solutions to differential
equations, are derived from recurrence relations, or from a generating 
function. In our case, since there are no clear constraints of this kind, 
we decided to adopt a very simple construction rule. 
Additional functions are generated by the parent function derivatives, 
as 
\begin{equation}
f_{n}^{m}\left(x\right)=\frac{d}{dx}f_{n-1}^{m}\left(x\right) = 
\left(\frac{d}{dx}\right)^{n}f_{0}^{m}\left(x\right)\,.
\label{eq:Base_Functions}
\end{equation}
 The overall set of functions will be addressed in the following as
``basis functions'', although a rigorous mathematical framework
supporting the term will not be implemented. It is thus considered
as just an expedient naming convention.

In the following, the focal plane units will either be micrometers
($\mu m$), pixels (1 pixel = 10 $\mu m$), or milli-arcsec (mas),
taking into account the Gaia aperture $L_{\xi}=1.45\, m$, the \emph{effective
focal length} of the telescope ($EFL=35\, m$) and the corresponding
\emph{optical plate scale} ($s=5.89\, arcsec/mm$).

The rationale leading us to test this particular approach is that
the signals of interest are expected to be reasonably close to the
ideal case, i.e. that they fit a context of small perturbation / small
aberration. Therefore, an expansion in terms of the aberration free
signal and related functions appears to be a promising
tool. Notably, even in case of large aberrations, as images for conventional
ground-based telescopes, the individual speckles are still described
by a superposition of displaced copies of the aberration free telescope
response, and the seeing image derives from integration of subsequent
speckles. The parent function takes advantage of one of the basic
aspects of the Gaia instrument, i.e. the telescope pupil size in the
main measurement direction (hereafter also high resolution
or along scan direction). 
Other basic factors of the instrument geometry,
characteristics and operation may be included in the basis functions,
as described below.

One peculiar aspect of the investigated basis functions is that they
are all referred, by construction, to a common ``zero point''
of the coordinates, corresponding (as centre of the aberration free
image) to the ideal position of the source image as provided by the
geometric optics. Also, they have simple symmetry: odd numbered functions,
as odd order derivatives of the even parent function, are odd, whereas
even numbered functions are even. In the numerical implementation,
since standard Matlab arrays or matrices are numbered from one onward,
the parity and function numbering are exchanged, i.e. the parent function 
is term no. 1, and so on. 
The lowest order
polychromatic functions, including the detection effects, are shown
in Fig. \ref{fig:Polychromatic-detected-LSF}.

\subsection{Polychromatic basis functions }

For any wavelength and position, it is possible to compute the monochromatic 
basis functions above. 
The parent function is defined by the geometry of the
ideal instrument; it can be computed, with its derivatives, in either
numerical or analytical form, using the trigonometry related expressions
from \ref{eq:Parent_function}, or the corresponding power series
expansion.

The superposition of monochromatic LSF terms at different wavelength
is weighted by the source spectral distribution, composed
with the instrument transmission distribution and the detector response
curve. The monochromatic basis is, by construction, source independent;
the polychromatic LSF construction factors out explicitly the
contributions from astrophysics (source spectrum) and astronomy (e.g.
reddening). The polychromatic LSF, and its derivatives, labelled as
$\{f_{0},\ldots,f_{N}\}$, must be computed numerically because of
the arbitrary weighting function corresponding to the effective 
spectrum $S(\lambda)$.
The polychromatic parent function can thus be expressed as \begin{equation}
f_{o}(x)=\int d\lambda\, S(\lambda)\cdot f_{0}(x;\lambda)\,,\label{eq:Poly_Parent}\end{equation}
 where the wavelength dependence of the monochromatic parent function
is explicited in Eq. \ref{eq:Parent_function}. The construction of
additional basis functions is straightforward, as from Eq. \ref{eq:Base_Functions}.

\begin{figure}
\includegraphics[scale=0.65]{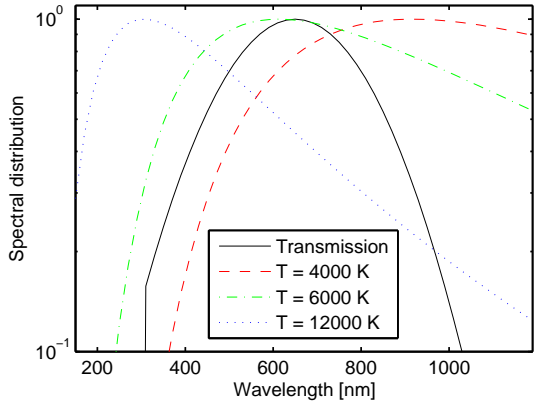} 
\caption{\label{fig:Spectral-distr}Spectral distributions of blackbodies 
at different temperatures superposed to the 
astrometric instrument response }
\end{figure}

We adopt a simple blackbody model for the source spectrum, which is
not a detailed representation of many astrophysical objects, but is
adequate to cover a realistic range of stars with respect to broadband
imaging; a representation of the Gaia spectral response and of the
normalised blackbody curves for three source temperatures 
is shown in Fig. \ref{fig:Spectral-distr}.%

\subsection{Detection effects \label{sub:Detection-effects}}

Additional modifications to the signal profile are induced by other
known parts of the detection process, in particular the geometric
effects of finite pixels, inducing a smoothing of the optical profile
through a rectangular filter with width corresponding to the pixel
size; the detector Modulation Transfer Function (MTF); the dynamical
mismatch between optical image and pixel array due to the Time Delay
Integration (TDI) operation. Some of the effects of finite sampling
and pixel size have been discussed, also in terms of the location
algorithm performance, in \citet{1998PASP..110..848G}. In the current
simulation, such contributions have been introduced as a wavelength
independent signal smoothing with realistic equivalent length,
respectively $10\,\mu m$ (geometric pixel size); $5\,\mu m$ (MTF);
$5.1\,\mu m$ (TDI). A more realistic wavelength dependent description
might in future be introduced e.g. for the MTF.

The superposition of wavelength contributions tends to average out
the function oscillations at increasing distance from the central
point. A sort of ``ortho-normalisation'' (depending on the current
sampling, i.e. offset) is applied, so that the integral of the product
of two basis function $f_{p},\, f_{q}$, over the detection interval
$\{x_{m}\}$, vanishes for different terms and is unity for the 
``diagonal'' term $p=q$: $\sum_{m}f_{p}\left(x_{m}\right)f_{q}\left(x_{m}\right)=\delta_{pq}$,
using the Kronecker's $\delta$ notation. An example of the resulting
set of functions is shown in Fig. \ref{fig:Polychromatic-detected-LSF}
for a $T_{s}=6000\, K$ source.

Remark: only the symmetric detection effects (pixel geometry, MTF
and TDI) have been included in the template, in order to preserve
the function symmetry. The simulated signals can include any kind
of degradation effect, which will appear in terms of distribution 
of the fit coefficients. 

\begin{figure}
\includegraphics[scale=0.65]{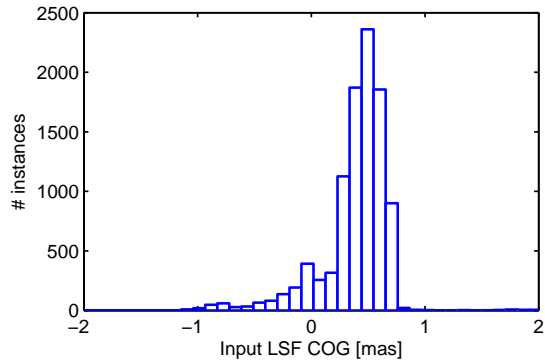} 
\caption{\label{fig:COG_hist}Histogram of the detected photo-centre 
values over the data set, zero detector offset }
\end{figure}

\section{Simulation}

\label{sec:Simulation} The goal of the fitting process is to reproduce
the aspects of interest of the measured signal $l(x_{k})$, corresponding
to the LSF generated by the optical system and detected by the CCD.
The sampled LSF is computed on a set of pixel centre positions $x_{k}$,
with $1\,\mu m$ resolution, i.e. 1/10 pixel, thus providing a high
resolution representation of the actual signal expected in operation.
The detected LSF $l$ is then fed to the fitting algorithm, deriving
the coefficients $c_{n}$ of a linear expansion referred to the basis
functions centred in a convenient location $\tilde{x}$: 

\begin{equation}
l(x_{k})=\sum_{n=1}^{N}c_{n}\cdot f_{n}(x_{k}-\widetilde{x})\,,\label{eq:Fit}\end{equation}
 where the equality is intended in the least square sense. The fit
quality can be evaluated in terms of consistency with the input data,
e.g. based on three criteria: 
\begin{itemize}
\item root mean square (RMS) discrepancy; 
\item integral difference (photometry); 
\item photocentre difference (astrometry). 
\end{itemize}
The first two items are strictly related, since two functions with
negligible RMS difference also have, at first order and under reasonable
assumptions, the same integral. Hereafter, the results will be discussed
based on the RMS discrepancy and photocentre difference only, using
for the latter the model independent barycentre or centre of gravity
(COG) algorithm. The COG is very simple, but in many respects not
practical for the Gaia data reduction with respect to both random
and systematic errors\citep{1978moas.coll..197L}.

\emph{It is assumed that a good fit, reproducing the input data profile
and position (with a given location algorithm), will also provide
consistent estimates of other parameters of interest, whichever the
selected algorithm.} Verification of performance and robustness of
specific algorithms should be considered in practical cases. 

The sample set for the main simulation includes 10000 different instances
of optical aberrations, generated by a random set of coefficients
for the Legendre polynomials of order up to 5 and 15, respectively
on the short (across scan) and long (along scan) side of the main
telescope aperture (respectively $0.5\times1.45\, m$), providing
a representation of wavefront error (WFE) down to the $0.1\, m$ scale.
Below, different signal instances are sometimes referred to by the
RMS WFE, i.e. the RMS value of the WFE over the telescope pupil; the
aberration free image has zero RMS WFE. The pupil is sampled with
resolution $1\, cm$ in both directions. The focal plane image resolution
is $1\,\mu m$ and $3\,\mu m$ respectively in the along and across
direction, corresponding in both cases to 1/10 of the geometric pixel
size. The spectral resolution is 20 nm, in the wavelength range 300
to 1100 nm; a realistic transmission curve is implemented by a Gaussian
distribution with $\sigma=250\, nm$ and peak at 650~nm. The focal
plane image is built by numerical computation of the diffraction integral,
according to the prescriptions in \citet{2007MNRAS.377.1337G}, and
integrated in the low resolution direction to represent the operating
mode of Gaia over a large fraction of the science data. The LSF is
computed with $1\,\mu m$ resolution over a region of $150\,\mu m$
(15 pixels), and the detection effects are included, as for the basis
functions. Representative readout samples corresponding to 12 pixels
(following nominal Gaia operation for intermediate magnitude objects),
with selected offset and $1\,\mu m$ resolution, are then used in
the fitting process described below.

For each instance, a different WFE and source temperature is used,
to cover a realistic range of variation of instrument parameters and
observed target. The source is represented by a simple blackbody distribution,
filtered by the instrument throughput (Fig. \ref{fig:Spectral-distr}).
In practical applications, realistic spectra can be inserted 
e.g. by means of data tables (relative intensity vs. wavelength) 
or other convenient descriptions. 

Remark: the nominal instrument configuration is associated to a limited
range of variation of the aberration coefficients describing the change
in optical response over the limited region of the focal plane used
by the detector. \emph{The simulation adopts a wide range of aberration
coefficients in order to cover not only the nominal values of the
relevant parameters of the optical configuration and of the detector
electro-optical characteristics, but also realistic modifications
of the in-flight system due to the transfer from ground to orbit
and consequent re-alignment. Also, limited degradation of such parameters
during operation, e.g. related to ageing or radiation damage, modifying
the detected signal profile, can described up to a point by an appropriate
change in the aberration coefficients. }

Therefore, our investigation provides indications on the capability
of the proposed fitting approach to follow the instrument response
evolution, of course by update of the coefficients through a convenient
calibration procedure. It is assumed that the system remains stable
over time periods sufficient for determination of the describing parameters
with sufficient precision and reliability. 
\begin{figure}
\includegraphics[scale=0.55]{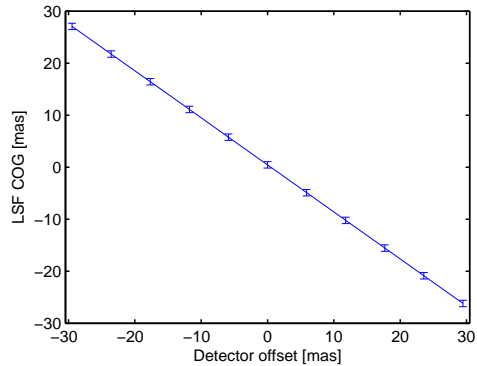} 
\caption{\label{fig:COG_hist_all} Average COG vs. offset, with RMS spread 
shown as an error bar }
\end{figure}

The
distribution of aberration instances generates a range of photocentre
values, evaluated on the zero offset sampled LSF by the COG algorithm,
and shown in Fig. \ref{fig:COG_hist}. Notably, the typical photocentre
displacement is below $1\, mas$, i.e. small with respect to the RMS
size of the LSF ($\sim130\, mas$), but significantly larger than
the measurement precision goal for intermediate magnitude stars (order
of $0.1\, mas$ at the elementary exposure level, and order of $10\,\mu as$
for the final catalogue). The spread in values associated to the statistical
sample has mean 0.446 mas, and RMS 0.619 mas. In this case, the optical
image is set with the coordinate origin coincident with the centre
of the detector pixel array, so that the aberration free image is
centred. The COG variation is due to distortion (in the strictly optical
sense) and to all other aberrations and degradation effects inducing
modifications in the signal profile, i.e. both actual translations
and deformations. Since the signal is sampled over a finite region
of the focal plane, the different truncation of an asymmetric, displaced
distribution affects both the photocentre and the fraction of energy
actually detected, i.e. the photometry.
\begin{figure*}
\includegraphics[scale=0.55]{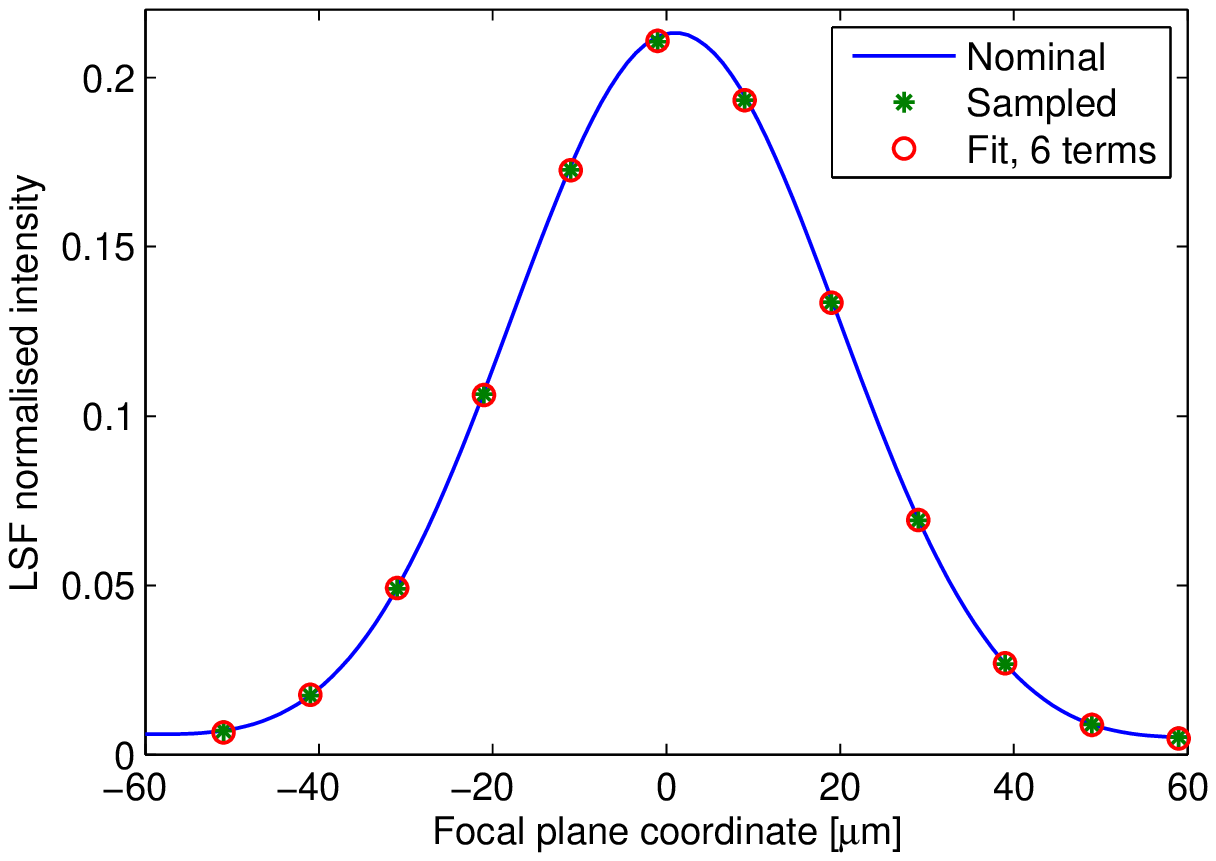} \includegraphics[scale=0.55]{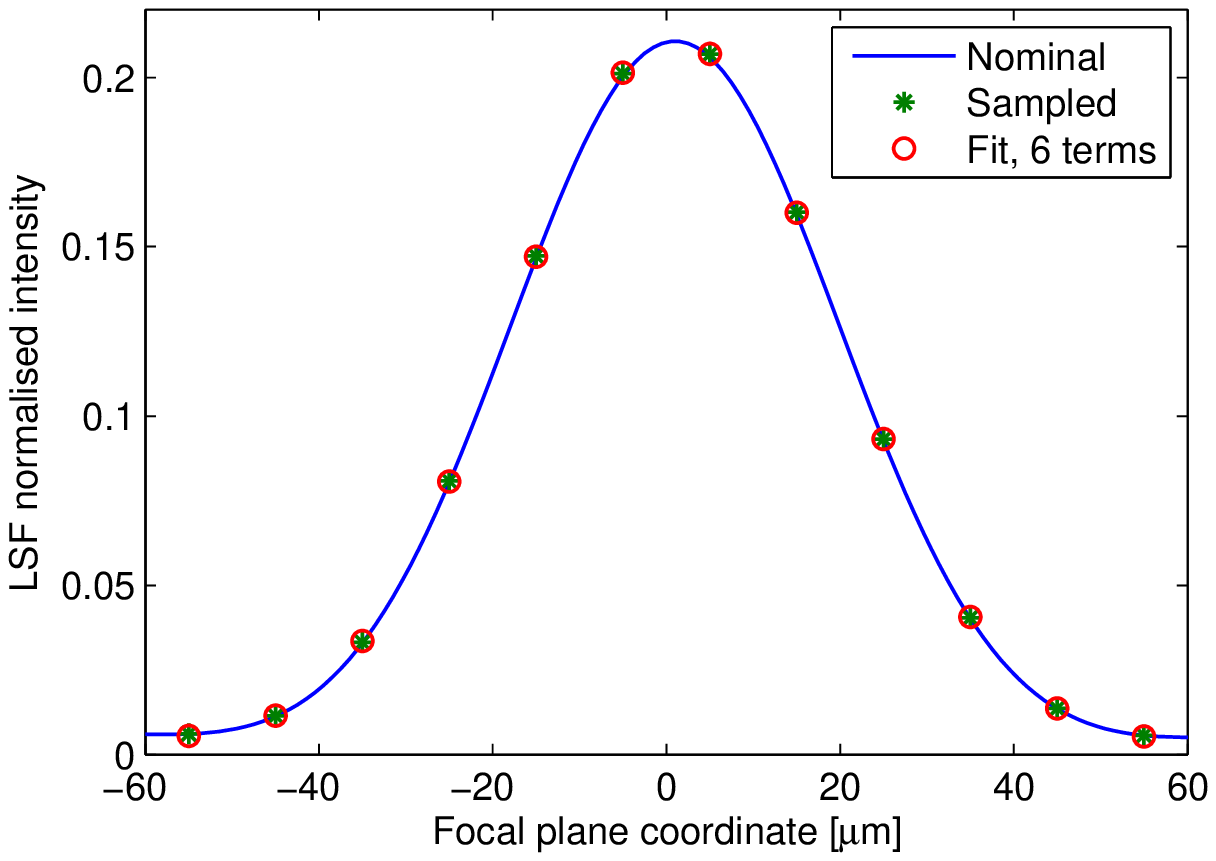} 
\caption{\label{fig:Examp-sampledLSF}Example of 12 pixel sampled LSF (crosses)
and its fit to 6 terms (circles), on top of the 1~$\mu m$ resolution
LSF (solid line), for offset zero (left) and 0.4 pixels (right); the
detected signal, i.e. relative pixel intensity, is a function of the
sampling offset }
\end{figure*}

Throughout the simulation, we apply a set of offsets to the LSF sampling
process, to represent a realistic range of displacement between the
optical image and the detector. A displacement larger than 0.5 pixel
just means that a different central pixel should be selected. The
range $\pm0.5$ pixel is covered with resolution 0.1 pixel. The overall
distribution of the sampled LSF COG, for all cases of offset, is shown
in Fig. \ref{fig:COG_hist_all}, evidencing that the COG spread remains
small for any applied offset, confirming
that the selected readout window is not affected by large signal variations.
The COG estimated on the detected image seems therefore to be a reasonable
first approximation for estimation of the detector offset, or correspondingly
for the actual image position vs. the detector considered as a reference,
and as such it will be used in the following steps of the simulation.

Notably, the correlation between detector offset and average COG,
shown in Fig. \ref{fig:COG_hist_all} (the error bars are the RMS
values of the COG distribution for each offset), is negative, since
a given offset applied to the detector corresponds to an image displacement
in the opposite direction. The estimated COG is not equal (in absolute
value) to the detector offset, because the LSF sampling positions
change: one of the LSF wings is pushed inside the sampling region,
the other outside, thus inducing a residual COG displacement. 
Typical residual values are below 10\% of the applied offset, and
comparable with the COG spread among different LSF instances, so that
the COG of sampled data provides a reasonable estimate of the mismatch
between LSF location and pixel array, and might be used as starting
approximation for more advanced algorithms. 

\subsection{\label{sub:AstrPhotSim}Astrometric and photometric fit}
For any WFE case, a set of offset values is introduced between the
pixel array and the sampling positions of the LSF, in the range $\pm0.5$
pixel, with resolution 0.1 pixel. 
The signal range corresponds to the 12 sample readout region, but with 
$1\,\mu m$ resolution, i.e. for the zero offset case the sampling positions 
are $[-55;\, -54;\, -53;\, \ldots;\, 53;\, 54;\, 55]\,\mu m$.  
The 110 points signal avoids any risk of fit degeneration using up to 
11 terms. 
The offset cases correspond to displaced sampling positions by 
$[\pm 1;\, \pm 2;\, \pm 3; \, \pm 4;\, \pm 5]\,\mu m$.  
The COG of the offset, sampled LSF 
is selected as reference point (origin)
of the basis functions used for signal fitting. 
In Fig. \ref{fig:Examp-sampledLSF},
a selected LSF instance (nr. 1) is shown, superposed to the 12 sample
LSF expected in operation (crosses) and to the corresponding fit result
using 6 terms from the basis function set (circles), respectively
for offset zero (left panel) and 0.4 pixel (right panel); the fitting
error is barely perceivable on this scale on the sides of the central
lobe. 

The best fit is computed against an increasing number of basis function
terms, to evaluate the most convenient number of terms required for
proper description of the sampled data derived from the input LSF.
The fit performance is then discussed.

\subsubsection{Fit residuals vs. offset }

\begin{figure*}
\includegraphics[scale=0.65]{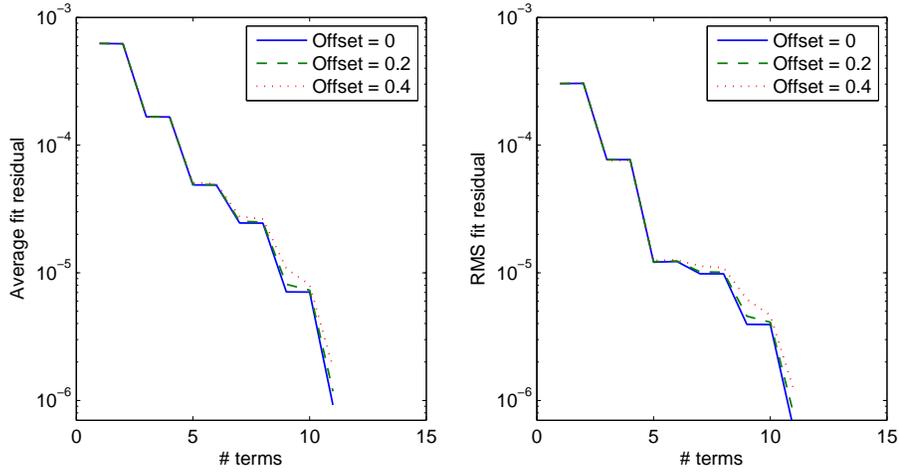} 
\caption{\label{fig:Fit-resid_11_0}Fit discrepancy RMS vs. number of 
terms up to 11, zero offset; average (left) and RMS (right) over 
the data set}
\end{figure*}

The RMS fit discrepancy for a few offset cases are shown in
Fig. \ref{fig:Fit-resid_11_0}, as a function of increasing number
of fitting terms. In particular, the left panel shows the average
over the sample (i.e. for different aberrations) of the RMS discrepancy
between the sampled LSF and the fit, as a function of the number of
basis functions used for the detected signal expansion. 
The RMS over the sample is shown on the right. 

The fit retrieves most of the input signal with 10 to 11 terms, according
to the progressively decreasing RMS over the
data set of the residuals with increasing number of fitting terms,
describing an increasing capability of the fit in capturing the fine
details of the LSF. The 11 terms case evidences that the fit is not
exact, but it provides a dramatic improvement with respect to 
lower dimensionality cases. 

Using one or two terms, the fit already accounts for more than 99\%
of the LSF energy, and with three or four terms the RMS discrepancy
decreases to about $10^{-4}$. Using five to eight terms, the RMS
discrepancy further drops to the $10^{-5}$ level, dropping to the
$10^{-6}$ level with 11 terms. %

\begin{figure*}
\includegraphics[scale=0.65]{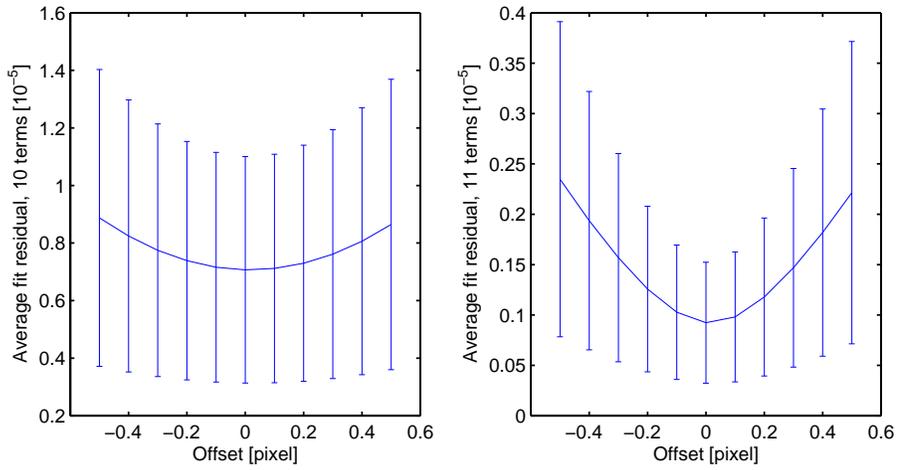} 
\caption{\label{fig:Fit_resid_10_11_off}Fit discrepancy mean and RMS (as 
error bar) vs. offset, using 10 (left) and 11 (right) terms}
\end{figure*}

The fit RMS discrepancy can be evaluated as a function of the offset
between LSF and pixel array, and the results for 10 (left) and 11
(right) fitting terms are summarised in Fig. \ref{fig:Fit_resid_10_11_off},
where it appears that the residuals increase with the absolute value
of offset, are affected by a significant variation over the LSF sample,
and improve by a factor four by switching from 10 terms
(average $\sim8\times10^{-6}$) to 11 terms (average $\sim2\times10^{-6}$).

Therefore, photometry and the image profile are basically retrieved
by modeling the LSF with either 10 or 11 terms of the proposed basis
functions. This result is not yet conclusive, since the crucial parameter
under investigation is the astrometric performance, dealt with in
the next section.

\subsubsection{Astrometric residuals vs. offset }

\begin{figure*}
\includegraphics[scale=0.65]{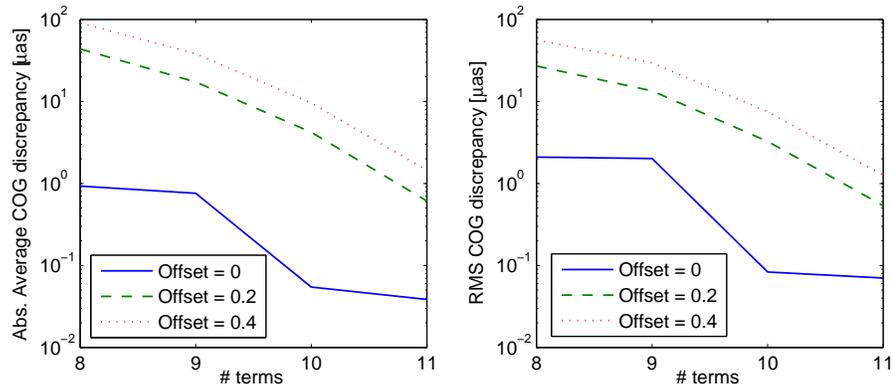} 
\caption{\label{fig:COG-discrepancy_0}COG discrepancy vs. number of 
terms from 8 to 11 for three cases of offset: absolute mean value (left) 
and RMS (right) over the data set }
\end{figure*}

Concerning the astrometric precision of the fit, Fig. \ref{fig:COG-discrepancy_0}
shows that, for the range 8 to 11 fitting terms, the COG of the reconstructed
LSF converges to the COG of the input sampled LSF, both as mean value
(left panel) and as RMS (right panel) over the data set, at increasing
number of terms. This is consistent with the results of the previous
section. %

\begin{figure*}
\includegraphics[scale=0.65]{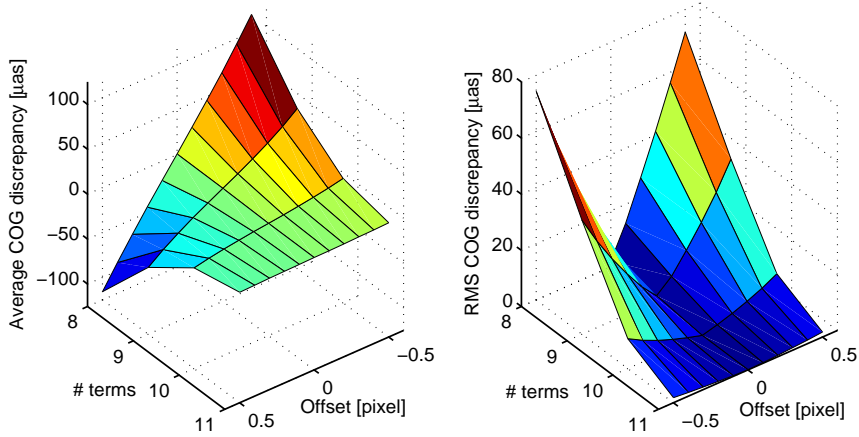} 
\caption{\label{fig:COG-RMS-offset}COG discrepancy vs. number of 
fitting functions and offset, mean (left) and RMS (right) values}
\end{figure*}
\begin{figure*}
\includegraphics[scale=0.65]{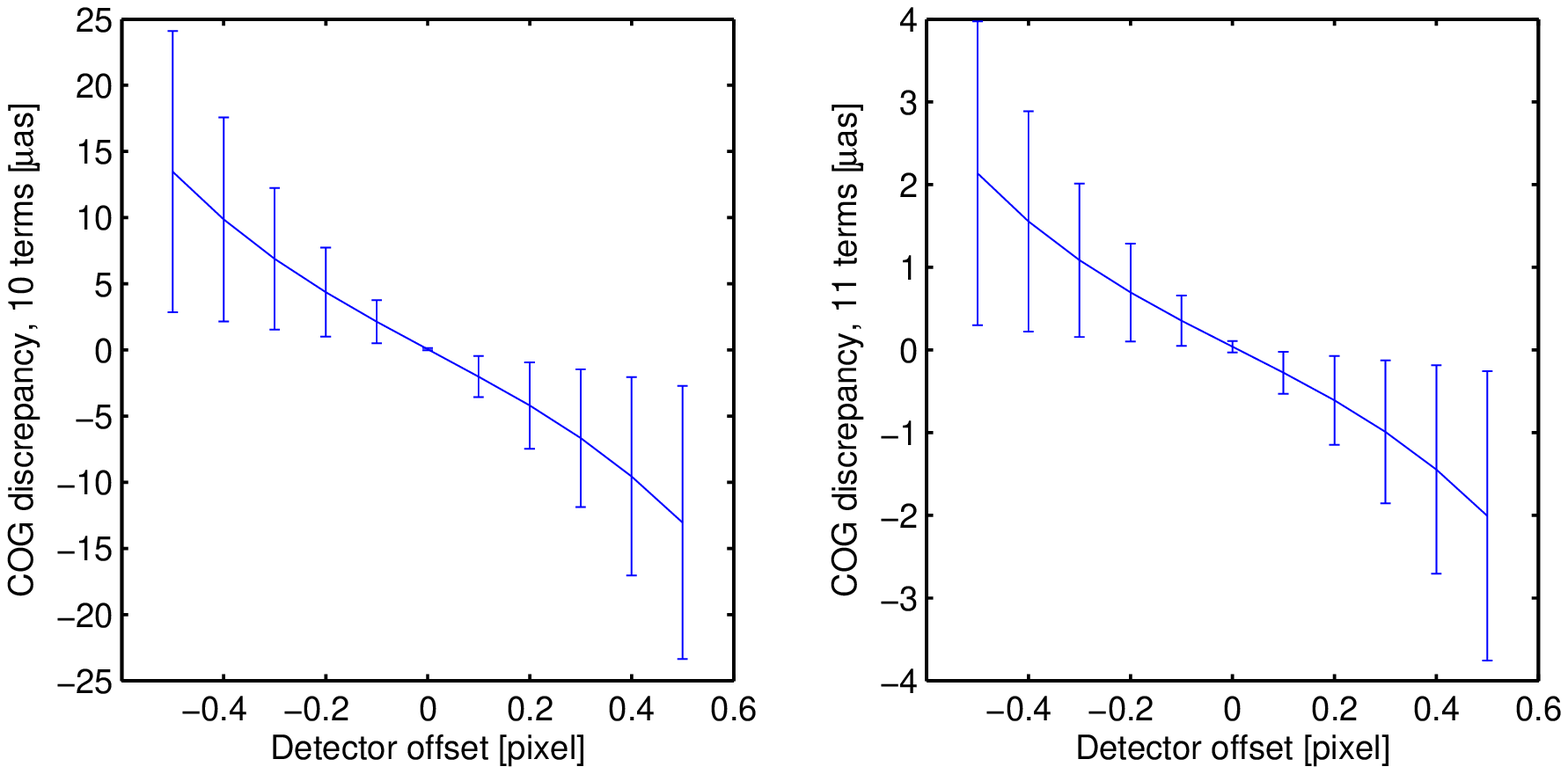} 
\caption{\label{fig:Mean-COG_10_11}Mean COG discrepancy vs. offset, 
for 10 (left) and 11 (right) fitting terms; RMS discrepancy shown 
as an error bar }
\end{figure*}

A COG error of few $\mu as$ RMS, around a comparable average error,
is achieved for the zero offset case with 8 terms. For larger offset
values, the discrepancy grows steeply, requiring either 10 or 11 terms
to retrieve $\mu as$ precision. 
The COG precision is shown as a function of the selected number of
fitting terms and of the offset between LSF and pixel array in Fig.
\ref{fig:COG-RMS-offset}, as average (left panel) and RMS (right
panel) distributions. The discrepancy is larger at increasing offset,
and decreases with increasing number of terms. Although most offset
cases are associated to larger RMS COG discrepancy than the zero offset,
for a given number of basis functions, convergence to the $\mu as$
level is still achieved when 11 terms are used.

The COG discrepancy is shown in Fig. \ref{fig:Mean-COG_10_11} respectively
for the case of 10 (left) and 11 (right) fitting terms, plotting the
mean value with a solid line and evidencing the RMS as an error bar.
The mean COG discrepancy remains within $1\sigma$ from the desired
zero value corresponding to unbiased signal reconstruction, but an
overall trend of the bias as a function of the pixel offset is present.
\\ 
We remark the dramatic improvement introduced by usage of 11 terms
rather than 10 or less, reducing the RMS and average COG discrepancy
from 15 $\mu as$ to 2 $\mu as$ peak, or from 7 $\mu as$ to 1 $\mu as$
on average vs. pixel offset. 
The scaling of centering residual is different from that of fit 
discrepancy (Fig. \ref{fig:Fit_resid_10_11_off}) because the location 
process is mostly sensitive to the steepest slope regions of the signal, 
thus affecting the error propagation 
\citep{1998PASP..110..848G}.

\subsubsection{Distribution of the fit coefficients }
\begin{figure}
\includegraphics[scale=0.6]{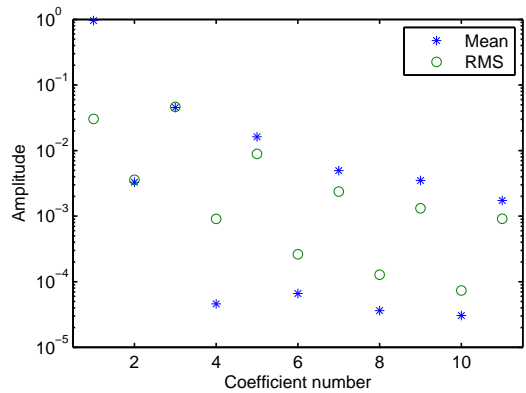} 
\caption{\label{fig:Ave-fit-coef_1}Fit coefficients for the zero offset case;
mean (stars) and RMS (circles) over the data set}
\end{figure}
The relative weight of the fit coefficients, up to 11 terms, for the
zero offset case is shown in Fig. \ref{fig:Ave-fit-coef_1}, as statistics
over the data set. 
\begin{figure*}
\includegraphics[scale=0.65]{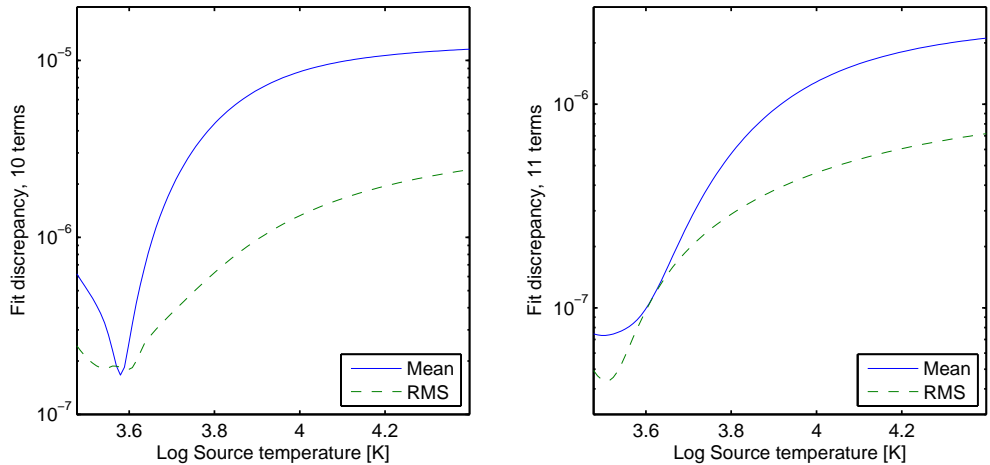} 
\caption{\label{fig:col_RMS-fit_10_11}Fit discrepancy vs. source 
temperature, 10 (left) and 11 (right) terms, zero offset; mean (solid line) 
and RMS (dashed line) over the data set}
\end{figure*}
\begin{figure*}
\includegraphics[scale=0.65]{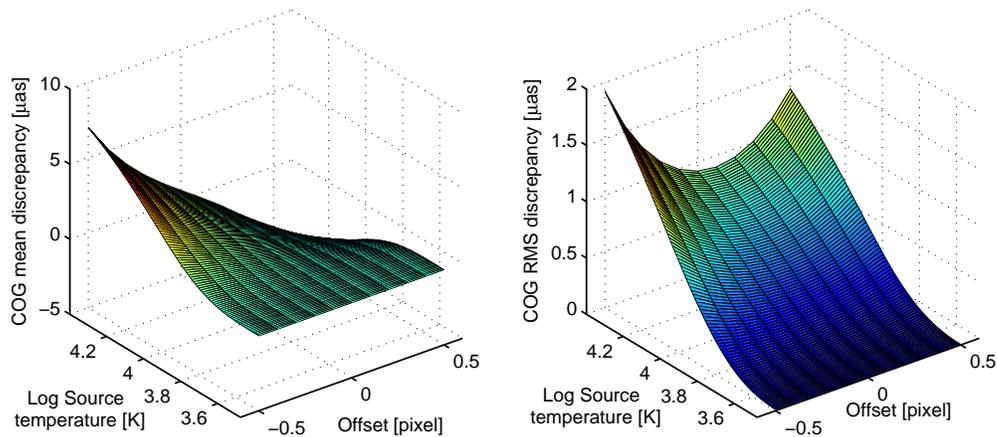} 
\caption{\label{fig:col_RMS_off}Average (left) and RMS (right) COG 
discrepancy vs. source temperature and offset, 11 terms}
\end{figure*}
The average (stars) is close to zero for even terms
(associated to odd functions), consistently with the expectations
for averaging over a representative random distribution of aberrations.
The RMS of the coefficients over the data set (circles) evidences
that their spread is also very small, i.e. that the individual values
are actually close to zero. The mean and RMS values for odd terms
(even functions) decrease for increasing order of the functions. This
is consistent with the simulation design of small aberration images,
with deviations from the diffraction limit mostly due to symmetric
degradation effects (pixel geometry, MTF, TDI).

\subsection{\label{sub:SourceTDep}Source temperature dependence}
The fit quality is evaluated as a function of the source spectrum
by generating a data sample, for a limited number of WFE instances
(100), with blackbody source temperature spread uniformly, in logarithmic
units, between 3000~K and 25000~K. The sample has thus 10000 instances,
as in the simulation in Sec. \ref{sub:AstrPhotSim}, but the optical
response variation is smaller, whereas the spectral coverage is finer.
The sample is then processed similarly to the case described in Sec.
\ref{sub:AstrPhotSim}, limited to 10 and 11 fitting terms, over a
range $\pm0.5$ pixel, always with resolution 0.1 pixel ($1\,\mu m$).

The zero offset case is first evaluated. The average fit discrepancy
remains consistent with the previous results, as shown in Fig. \ref{fig:col_RMS-fit_10_11},
where the average (solid line) and RMS (dashed line) discrepancy are
shown in logarithmic units vs. the source temperature. 
The spread is associated to the different aberration instances. 
Using 10 terms, the average discrepancy
is below $10^{-6}$ for near-solar and later spectral types, and it
increases for earlier types to $\sim10^{-5}$. A similar trend is
achieved for the 11 term case (right panel), with values reduced by
roughly a factor five. 

\begin{figure*}
\includegraphics[scale=0.65]{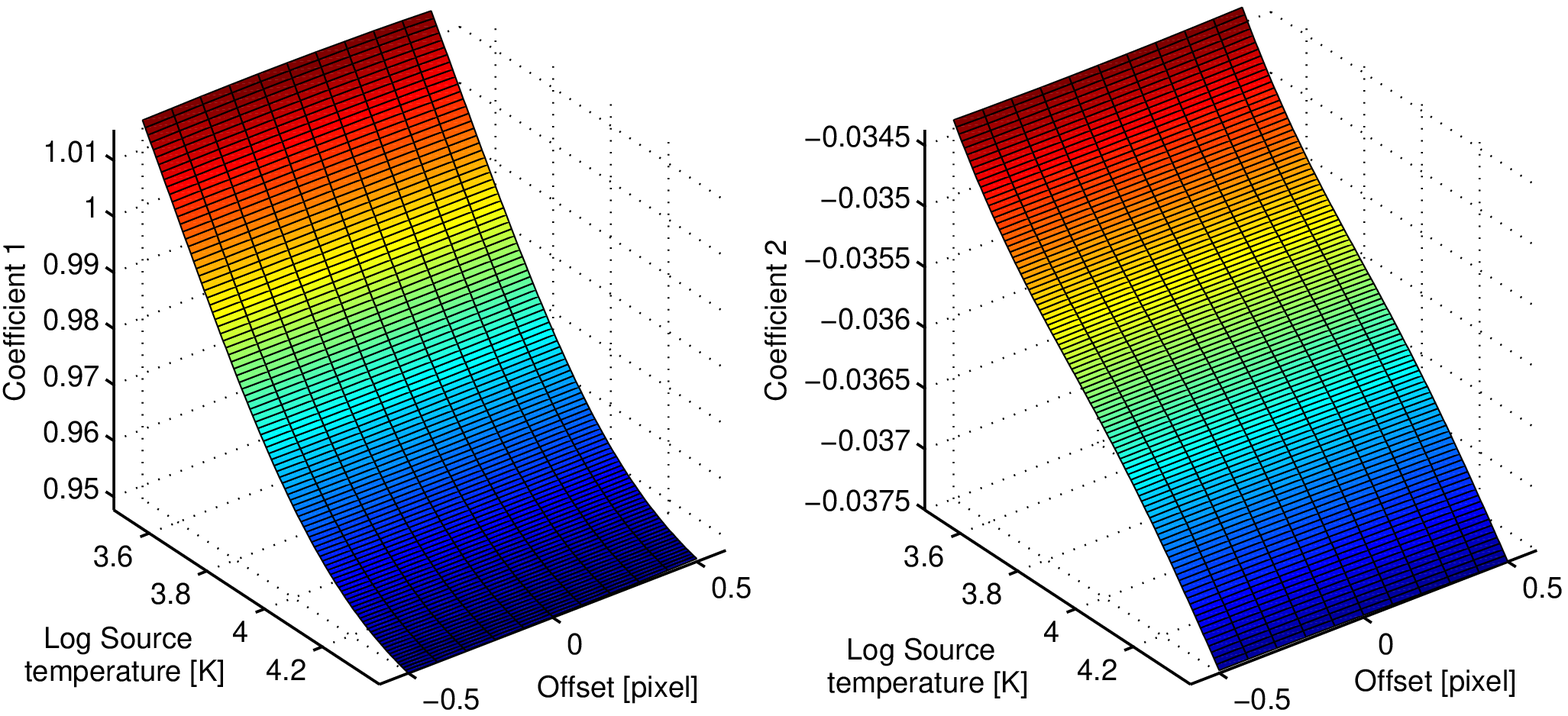} \includegraphics[scale=0.65]{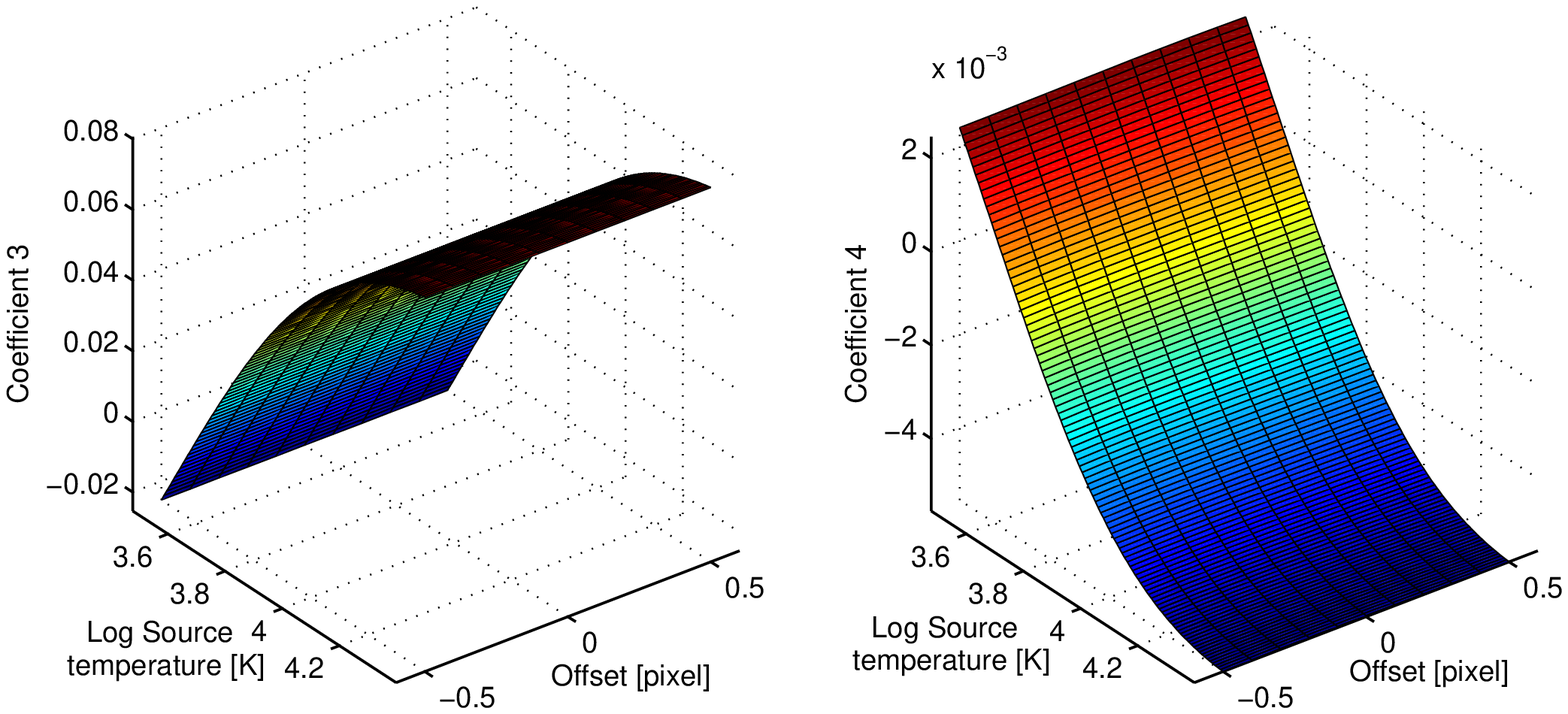} 
\caption{\label{fig:Fit_1_4_WFE1}Fit coefficients 1 to 4 vs. offset and source
temperature, WFE instance 1}
\end{figure*}
The dependence of astrometric discrepancy from offset can then be
considered. The average and RMS values vs. source temperature and
offset, 11 terms, are shown in Fig. \ref{fig:col_RMS_off}, left and
right panel respectively. The average discrepancy reaches few $\mu as$
peak values, whereas the RMS remains below $2\,\mu as$ peak; the
fit discrepancy is below $1\,\mu as$ for low temperature cases.
Using 10 terms, the discrepancy is somewhat degraded, respectively
to a few $\mu as$ RMS and a few 10 $\mu as$ on average over the
data set. The error increase with the source temperature, being very
small for near solar and later spectral types, will be further discussed
in Sec. \ref{sec:Calibration}. 
\\ 
The variation of the fit coefficients 1 to 4 for the WFE instance
no. 1, with both source temperature and offset, is shown in Fig. \ref{fig:Fit_1_4_WFE1}.

The variation of the fit coefficients over the data set depends on
both source spectral characteristics (e.g. the source temperature
or its logarithm, in the simple blackbody model) and sampling offset.
The dependence is very smooth, as for the COG discrepancy (Fig. \ref{fig:col_RMS_off}),
so that is appears that it could be expanded in the form of a very
simple function, e.g. low order polynomials. Besides, different instances
of aberrated images exhibit, as could have been expected, different
spectral dependence \citep{2006A&A...449..827B}. Therefore, it appears
that the coefficients may be mapped over the field of view by calibration
with comparable ease. 

\subsection{\label{sub:Tol_sourceT}Sensitivity to source temperature}

In the simulation described in Sec. \ref{sub:AstrPhotSim}, the random
variation of both aberration and source temperature does not allow
to evidence simple trends as those shown in Sec. \ref{sub:SourceTDep}.
Besides, the simple dependence on source temperature evidenced in
Sec. \ref{sub:SourceTDep} suggests to investigate directly on this
data set the sensitivity to errors in the knowledge of the source
temperature. The same input data is used, again with $1\,\mu m$ resolution
and ranging over $\pm0.5$ pixel offsets; however, in the current
run, a $\pm1\%$ error on the source temperature is applied in the
construction of the basis functions, thus representing an uncertainty
on the knowledge of the source temperature or a spectral distribution
variation either in the source itself (e.g. in case of a variable
star) or in the instrument response. The fit is performed using both
10 and 11 terms.
\begin{figure*}
\includegraphics[scale=0.65]{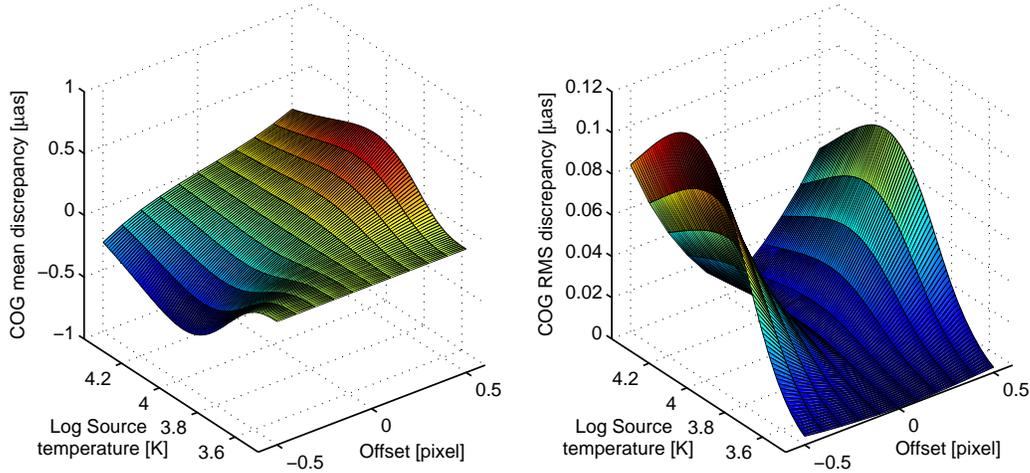} 
\caption{\label{fig:DCOG_DT10} Mean (left) and RMS (right) COG 
discrepancy vs. offset and source temperature, with temperature 
error, 10 terms  }
\end{figure*}
\begin{figure*}
\includegraphics[scale=0.65]{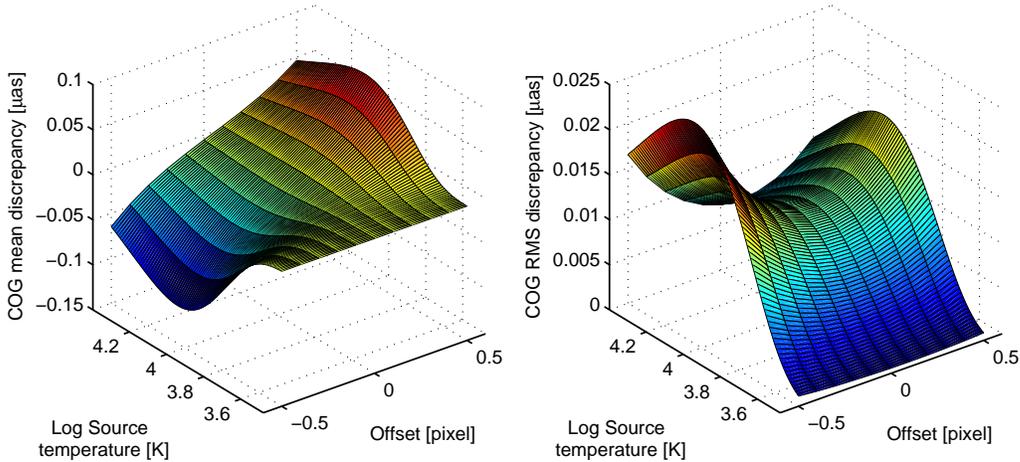} 
\caption{\label{fig:DCOG_DT11}Mean (left) and RMS (right) COG 
discrepancy vs. offset and source temperature, with temperature 
error, 11 terms  }
\end{figure*}

The COG estimates for both $+1\%$ and $-1\%$ source temperature
errors are then compared in order to assess the consequences on the
measurements. 

The COG difference between the two cases 
is shown vs. offset and temperature 
in Figs. \ref{fig:DCOG_DT10} and \ref{fig:DCOG_DT11}, respectively
for the case of 10 and 11 fitting terms, evidencing the mean (left)
and RMS (right) values over the data set. 
The COG difference of either
$+1\%$ or $-1\%$ error from the nominal source temperature COG result
is smaller (about half as much). 
Over a 2\%
variation of the source temperature, using 10 terms, the mean COG
discrepancy remains below $1\,\mu as$, with values much smaller for
either low source temperature or small offset, and some mitigation
also for very high temperature values. Correspondingly, the RMS COG
discrepancy between the two cases is below $0.1\,\mu as$, with peaks
corresponding to intermediate temperatures and large offset. Using
11 basis functions, the COG discrepancy is reduced by nearly one order
of magnitude both in terms of mean and RMS values. 

The astrometric error introduced by a $\pm1\%$ error on the knowledge
of the source temperature, therefore, induces a marginal variation
($\sim1\%$) in the photocentre reconstruction, as seen
e.g. by comparison with Fig. \ref{fig:col_RMS_off} or Fig. \ref{fig:Mean-COG_10_11}.

Since the fit error is very small, in spite of the temperature
error, the sensitivity to knowledge of the source temperature,
or to related variations, is small, even for the 10 term case. Assuming
linear scaling, a 10\% error on the source temperature will remain
acceptable to a measurement accuracy of few $\mu as$ for most cases
of optical response, using either 10 or 11 terms for the fit. Additional
simulations may be required to provide better quantitative estimates
of the sensitivity, due to the limited number of aberration instances,
and model limitations at the sub-$\mu as$ level. 

\begin{figure*}
\includegraphics[scale=0.65]{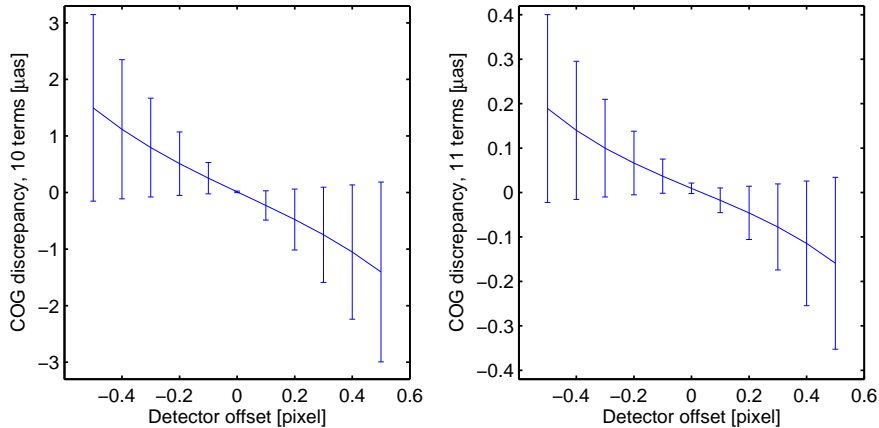} 
\caption{\label{fig:COG-discr_T65k}Mean COG discrepancy vs. offset (RMS as
error bar), for 10 (left) and 11 (right) fitting terms, for instances
associated to source temperature below 6500 K }
\end{figure*}

\section{Calibration aspects}

\label{sec:Calibration} The fit test discussed in Sec. \ref{sub:AstrPhotSim}
is somewhat artificial, since it is not applicable directly to science
data due to simulation dependence on 
\begin{enumerate}
\item pixel level sampling of real data, i.e. 10 $\mu m$ resolution rather 
than 1 $\mu m$; 
\item pixel offset, i.e. relative phase between the currently observed target
and the pixel array; 
\item source spectral distribution; 
\item photon limited information on individual exposures. 
\end{enumerate}
Actually, a model of the effective LSF should be derived from a set
of data corresponding to several objects observed at different pixel
offsets, as provided naturally by the spread of star positions on
the sky. 
This procedure must also account for the variation of the coefficients 
with the source spectral characteristics; the composition of data also 
improves on the photon limit issue. 
It is necessary to feed the fit with a sufficiently large astrophysical 
sample, since individual exposure data are to be weighted according to 
their statistical significance, i.e. SNR and source brightness. 
The most convenient approach for practical implementation in the Gaia
data reduction system should be further investigated. 
The above simulation approach was adopted for its simplicity, and
is deemed adequate for a first assessment of the relevant properties
of the proposed model, but it does not meet \emph{per se} 
all needs for implementation in the data reduction pipeline.

The offset cases are not representative of the complete LSF 
reconstruction process: they relate to the case of individual 
observations, with the aim of identifying potentially relevant 
contributions to the systematic error. 
A few instances of bright, hot objects, detected with large offset, 
might introduce a comparably large bias in the reconstructed LSF, 
significant with respect to their photon limited location precision. 
The simulation results suggest that the data reduction system should 
monitor the LSF reconstruction against this situation, possibly 
introducing corrections. 
Still, usage of enough terms in the LSF modelling (up to 11) 
significantly reduces the individual bias contribution. 

\begin{figure}
\includegraphics[scale=0.6]{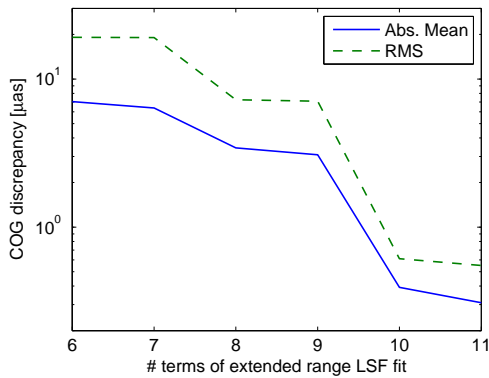} 
\caption{\label{fig:COG_ext}Extended range LSF model: astrometric 
discrepancy as a function of the number of fitting terms 
over $\pm 60 \, \mu m$ }
\end{figure}

As an idealised case of real data composition, we assume that a large 
number of single exposure data are collated to cover the whole offset 
range and used to fit a single, zero offset LSF model; the sampling 
positions are then set to 
$[-60;\, -59;\, -58;\, \ldots;\, 58;\, 59;\, 60]\,\mu m$, 
corresponding to 12 pixels (the readout region) $\pm 0.5$ pixels 
to account for the offset due to individual object positions. 
The issue of correct photocentre determination is neglected at the 
moment, as well as possible weighting based on offset, SNR and colour. 
The astrometric discrepancy vs. number of fitting terms of such high 
resolution LSF model with respect to the parent data set is shown in 
Fig. \ref{fig:COG_ext}. 
The fit quality is at the $\mu as$ level RMS with 10 terms, and the mean 
value is somewhat lower; in practical cases, this suggests that the fit 
is expected to have negligible systematic error and noise dominated by 
the available amount of photon limited exposures. 

The relevance of offset between optical image and detector
may be better appreciated in terms of the equivalent
total distortion, considered as the overall set of optical effects
inducing image displacement. From the point of view of signal profile
fitting, the applied displacement up to $\pm0.5$ pixel, i.e. $\pm29\, mas$,
corresponds to about $\pm1/6$ of the Airy diameter at $\lambda=600\, nm$
(about $30\,\mu m$, or 171~mas). Therefore, it is larger than the
overall optical aberration introduced in the simulation sample, with
average value of the RMS WFE below $\lambda/10$, and much larger
than the astrometric effect induced on the images, below $\lambda/100$
(Fig. \ref{fig:COG_hist}).

As shown in Sec. \ref{sub:SourceTDep}, in case of sources with
simple spectral distribution, known with an acceptable tolerance (Sec.
\ref{sub:Tol_sourceT}), a ``grid'' of calibration instances covering
e.g. 10 to 20 different temperatures over the desired range, and order
of 10 different offsets on the detector, is expected to provide satisfactory
results. 
Besides, simple SNR considerations lead to the need of averaging
many individual measurements in order to achieve photon driven
precision comparable with the desired fit precision. 
For example,
setting a goal of $10^{-5}$ on the RMS fit discrepancy corresponds
to a requirement on the required cumulative SNR $\sim10^{5}$, independent 
of the selected LSF expansion strategy. 
This corresponds to matching
the fit precision associated to a given number of terms with the photon
limit on the knowledge of the effective signal profile. Therefore,
a 10 term fitting model, with lower intrinsic precision ($10^{-5}$),
has more relaxed calibration requirements, since it is much easier
(or faster) to accumulate the corresponding amount of photon limited 
data ($SNR\simeq10^{5}$).
Model monitoring procedures at the $10^{-5}$ level can then 
be applied on a time scale (i.e. data amounts) shorter by a factor
100 with respect to similar procedures aimed at the $10^{-6}$ precision
goal, roughly corresponding to 11 fitting terms. 

Also, taking advantage of the fit quality vs. source temperature (Figs.
\ref{fig:col_RMS-fit_10_11} and \ref{fig:col_RMS_off}), it might
be found convenient to split the model into a part referred to near
solar type objects, and a part with additional chromatic corrections
for other objects. For example, using the data from Sec. \ref{sub:AstrPhotSim}
and restricting the data to lower source temperature ($T_{s}<6500\, K$),
corresponding to about 3000 instances over 10000, the distribution
of COG discrepancy vs. offset is shown in Fig. \ref{fig:COG-discr_T65k}.
By comparison with the corresponding Fig. \ref{fig:Mean-COG_10_11},
referred to the whole data set, a precision improvement by about one
order of magnitude is achieved for both cases of either 10 (left)
and 11 (right) fitting terms. 
In this case, $\mu as$ level precision is achieved already with 10 terms.

The better fit quality for near-solar spectral types can be related
to the selected spectral passband of Gaia (Fig. \ref{fig:Spectral-distr}),
which collects a significant fraction of their blackbody distribution
and includes their maximum. Warmer stars have a large fraction of
their blackbody distribution outside the Gaia observing band, so that
a comparably small change in the source temperature induces a significant
displacement in the detected spectrum and its effective wavelength.
A similar consideration holds for colder stars as well, but the image
is in any case less affected by aberrations at longer wavelengths,
due to the WFE scaling in the diffraction
integral as $WFE/\lambda$.

\section{Discussion}

\label{sec:Discussion} The simulation of Sec. \ref{sub:AstrPhotSim}
adopts a fitting strategy in which the basis function center is initially
set to the sampled LSF COG, and never modified; only the function
coefficients are adjusted to achieve the best fit, in the least square
sense, with the input data. Since the goal is an LSF model reproducing
both data profile \emph{and} location, the correct approach would
require that, for a selected number $N$ of fitting terms, a complete
set of $N+1$ parameters were estimated for best fit with the data,
i.e. the $N$ function coefficients and the additional term defining
the new location estimate. \emph{Therefore, the fitting and location
algorithms become entangled.} 

Moreover, due to the form
of the basis functions, the model is no longer linear in its parameters,
and the conventional least square approach cannot be considered to
be mathematically correct. However, due to the assumptions of small
deviations from the aberration free signal, the correlation between 
photocentre
and fit coefficients may be comparably loose, and
an iterative solution can be expected to converge for all parameters.
In particular, the function coefficients could be approximated by
setting the initial values as $c_{0}=1$; $c_{2}=\ldots=c_{N}=0$.

This approach was not investigated in this stage of development, under
the assumption that the simplest, location independent fitting algorithm
already provides sufficient indications on the achievable fitting
performance of the proposed model. The fit convergence in terms of
rapidly diminishing values of both RMS discrepancy and COG difference
seems to confirm the soundness of the proposed strategy. More complex
fitting approaches will be considered in future investigations. %
\begin{figure*}
\includegraphics[scale=0.6]{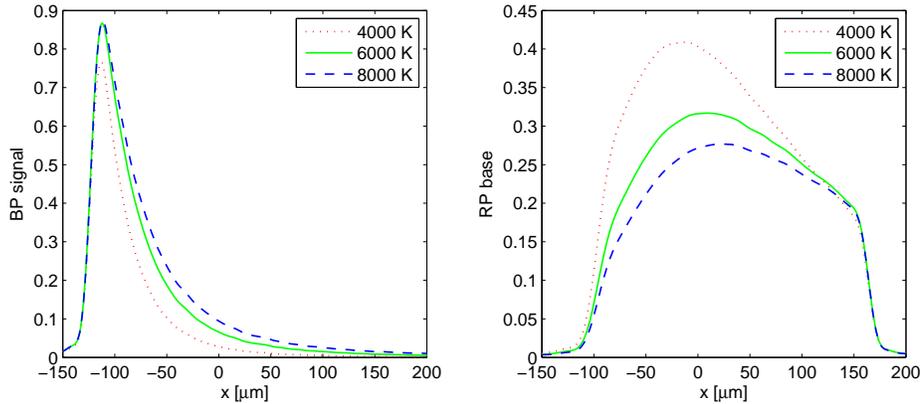} 
\caption{\label{fig:Spectral-distr-XP}Parent function for the BP (left) and
RP (right) instruments for different source temperatures}
\end{figure*}

The model fitting was implemented over a limited readout region (12
pixels) to match the nominal Gaia operation on intermediate brightness
objects; the extension to faint objects, on which a lower number of
readout samples is extracted, and to a number of other operation modes,
is conceptually straightforward. The bright end must be considered
explicitly, since images (conventionally called PSF, with a loose
extension of the standard optical definition of the Point Spread Function
term) are read as bi-dimensional windows, without binning in the low
resolution direction. The bi-dimensional fit might, in principle,
require different functions or, at least, in the simplest extension,
more parameters to generate ``2D'' basis functions by composition
of their one-dimensional counterparts. 
However, the across scan resolution on both pupil and focal plane is 
smaller, as well as the goal measurement precision, and for small 
aberrations it can be expected that sufficient precision could be 
achieved by using a limited number of terms. 
The subject will be addressed in future studies. 

The LSF fit investigated in this study was focused on the central
lobe of the LSF, addressing the science data modelling performance.
Besides, the basis functions can be computed at arbitrary distance
from their centre, so that they can be used conveniently also for
representation of the LSF wings, e.g. at some distance from bright
objects, to investigate the contamination on other stars. The limitations
on this subject are not imposed by the model, but rather by the limited
knowledge on the related instrument parameters (realistic values of
high spatial frequency manufacturing errors, micro-roughness, dust
contamination etc.).

The basis function definition may be modified, in order to improve
on specific properties, e.g. to reduce the parameter dependence on
astrophysical source variation by adopting different spectral weighting
of the monochromatic functions, or to ease the numerical implementation
from the standpoint of processing, robustness and other relevant aspects.
The range of some aberrations might be restricted, since perturbations
of realistic configurations often do not change the WFE shape in the
same way for all describing parameters (e.g. Legendre or Zernike coefficients).
The reduced range could thus be sampled with higher resolution, thus
providing more reliable and detailed results. Conversely, larger aberrations
may require additional fitting terms in order to retain $\mu as$
level precision. A series of simulations is planned for exploration
of several such options.

The proposed signal model can be applied, with straightforward modifications,
to the photometric and spectroscopic sections of the instrument. In
this case, the polychromatic signal (taking place of the polychromatic
parent function in Sec. \ref{sec:Signal_model}) is still built as
a superposition of the monochromatic terms, and they are no longer
referred to the same focal plane positions, but rather affected by
a displacement related to the appropriate spectral dispersion. In
Fig. \ref{fig:Spectral-distr-XP}, a representation of the parent
function for the two photometric channels of Gaia, labelled Blue and
Red Photometers, resp. BP (left) and RP (right), is shown for three
values of source temperature, using a simple dispersion law. The parent
functions represent the ideal instrument response; the derivatives
build also in this case the additional basis functions which can be
used to fit the realistic signal.

The application of the basis model to conventional circular pupil
telescopes is straightforward, by replacement of the sine 
in the parent 
function (Eq. \ref{eq:Parent_function}) with the Bessel function
$J_{1}(\rho)$ appropriate to the geometry.

\section*{\label{sec:Conclusions}Conclusions}

The LSF representation for the astrometric field of Gaia is addressed
by means of a set of functions based on the aberration free response
of the ideal telescope and its derivatives, composed according to
the source spectral distribution. The simulation takes into account
the instrument response variation as a function of the relative position
of the detector pixel array with respect to the optical image, evaluating
its effect on the model parameter estimation. The fit quality is evaluated
as a function of the RMS discrepancy and photocentre difference with
the input data; both criteria result in error drops with increasing
number of fitting terms, down to negligible values (respectively below
$10^{-5}$ and $1\,\mu as$ RMS) by using 11 terms.

The calibration of the fit parameters on science data is straightforward,
based on sets of observations spannig a convenient range of different
spectral types and observing offset with respect to the detector geometry,
as implicitly provided by the natural distribution of stars over a
significant fraction of the sky. This approach also provides the necessary
averaging of photon noise. 
The requirements on astrophysical parameters of individual sources, 
correspond to of order of 10\% on the effective temperature. 

Possible improvements on the understanding of the properties of the
proposed fitting model, by more detailed simulation and evolution
of the model, are discussed, also suggesting other applications, basis
function modifications (also to include bi-dimensional signal modelling)
and implementation upgrades. Future investigations are planned on
several of the above issues.

\section*{Acknowledgments}

The study presented in this paper benefits from the discussions with
colleagues in the Gaia Data Processing and Analysis Consortium (DPAC),
and in particular with F. Van Leeuwen, L. Lindegren and M.G. Lattanzi.
The activity is partially supported by the contracts COFIS and ASI 
I/037/08/0. 

\bibliographystyle{mn2e} 
\bibliography{MyBibList}


\label{lastpage}
\end{document}